\documentclass[runningheads]{Template/llncs}

\usepackage{graphicx}
\usepackage{framed}
\usepackage{multirow}
\usepackage{xcolor}
\usepackage{booktabs}
\usepackage{tabularx}
\usepackage{siunitx}
\usepackage{amsmath}

\usepackage{hyperref} 

\definecolor{forestgreen}{rgb}{0.13, 0.54, 0.13}

\DeclareMathOperator{\loss}{loss}

\newcommand{\ie}{\textit{i}.\textit{e}.~}

\newcommand{\etal}{\textit{et~al.~}}

\usepackage{todonotes}

\graphicspath{{Images/}}

\begin{document}
\title{Multi-stage Deep Layer Aggregation for Brain Tumor Segmentation}

%%%%%%%%%%%%%%%%%%%%%%%%%%%%%%%%%%%%%%%%%%%%%% 
%%%%%%%%%%%%% List of authors %%%%%%%%%%%%%%%%
\author{
Carlos A. Silva\inst{1} \and
Adriano Pinto\inst{1} \and
Sérgio Pereira\inst{1} \and
Ana Lopes\inst{1}
}
%%%%%%%%%%%%%%%%%%%%%%%%%%%%%%%%%%%%%%%%%%%%%%
\authorrunning{C. Silva \etal}
\institute{
Center MEMS of University of Minho, Campus of Azurém, \\ 
4800-058 Guimarães Portugal \email{csilva@dei.uminho.pt} %\and
}

% show title
\maketitle
\begin{abstract}
    Gliomas are among the most aggressive and deadly brain tumors. This paper details the proposed Deep Neural Network architecture for brain tumor segmentation from Magnetic Resonance Images. The architecture consists of a cascade of three Deep Layer Aggregation neural networks, where each stage elaborates the response using the feature maps and the probabilities of the previous stage, and the MRI channels as inputs. The neuroimaging data are part of the publicly available Brain Tumor Segmentation (BraTS) 2020 challenge dataset, where we evaluated our proposal in the BraTS 2020 Validation and Test sets. In the Test set, the experimental results achieved a Dice score of $0.8858$, $0.8297$ and $0.7900$, with an Hausdorff Distance of $5.32$ mm, $22.32$ mm and $20.44$ mm for the whole tumor, core tumor and enhanced tumor, respectively.

\keywords{Brain tumor segmentation  \and Deep Learning \and Convolutional Neural Networks \and Gaussian filters.}
\end{abstract}
\section{Introduction}

Gliomas present the highest mortality rate and incidence among brain tumors. They can be categorized according to their aggressiveness into two levels: High Grade Gliomas (HGGs) and Low Grade Gliomas (LGGs) \cite{menze2014multimodal}. Currently, multi-sequence Magnetic Resonance Imaging (MRI) is the best imaging modality to assess the structure and sub-regions of gliomas, due to their highly heterogeneity in appearance, shape, location, and tissues. MRI allows the volumetric characterization of the lesions, which requires semantic segmentation. However, expert manual segmentation is expensive, susceptible to inter- and intra-rater variability, and time-consuming \cite{menze2014multimodal,bakas2018identifying}.

The heterogeneous nature of gliomas make automatic segmentation very challenging. So, in the past years, Convolutional Neural Networks (CNNs) that learn complex features directly from the data have achieved the best performances in the Brain Tumor Segmentation (BraTS) challenge \cite{pereira2016brain,myronenko20183d,jiang2019two,zhao2019bag,kamnitsas2017ensembles}. Pereira \etal \cite{pereira2016brain} proposed a point-wise segmentation approach using plain CNNs. Despite its success, following approaches \cite{myronenko20183d,jiang2019two,zhao2019bag,kamnitsas2017ensembles,pereira2019adaptive} were mostly based on more efficient Fully Convolutional Network (FCN) principles \cite{shelhamer2017fully}, especially inspired on the encoder-decoder U-Net architecture \cite{ronneberger2015u}. Kamnitsas \etal \cite{kamnitsas2017ensembles} ensembled several different models to tackle the bias-variance trade-off of the models. Myronenko \cite{myronenko20183d} proposed a 3D FCN network for gliomas segmentation, but coupled an extra decoder based on Variational Auto-encoder with the purpose of regularizing the encoder. Zhao \etal \cite{zhao2019bag}, as the authors define it, proposed a bag of tricks to enhance training and inference. Some of the procedures include careful voxel sampling, multi-size patches, extra unlabeled data for semi-supervised learning, and multi-task learning. Jian \etal \cite{jiang2019two} ranked 1\textsuperscript{st} in BraTS 2019 by developing a two-stage cascade FCN, where the second FCN is able to refine the results from the first. Inspired by \cite{myronenko20183d}, they also employ an extra regularization decoder with a simple interpolation scheme.

The U-Net \cite{ronneberger2015u} architecture includes long skip connections to merge low and high level features of the same scale. Therefore, these connections are shallow. Instead, the work on Deep Layer Aggregation (DLA) \cite{yu2018deep} proposes a principled iterative and hierarchical aggregation of layers from all scales. In this way, it enables the model to gather more spatial and semantic information for better segmentation refinement. In this paper, we evaluate the use of DLA blocks for brain tumor segmentation. To further refine the segmentation, we employ a three-stage cascade architecture, where the output of one FCN is directly fed to the next. This is an alternative to directly going deeper by down-sampling layers. The whole model is trained end-to-end.

The remaining of this paper is organized as follows. In Section \ref{sec:method}, we present the proposed method. The data and experimental setup are described in Section \ref{sec:setup}. Then, in Section \ref{sec:results}, the experimental results are presented, followed by the main conclusions in Section \ref{sec:conclusion}.

\section{Methods}
\label{sec:method}

\begin{figure}[!h]
		\includegraphics[width=\textwidth]{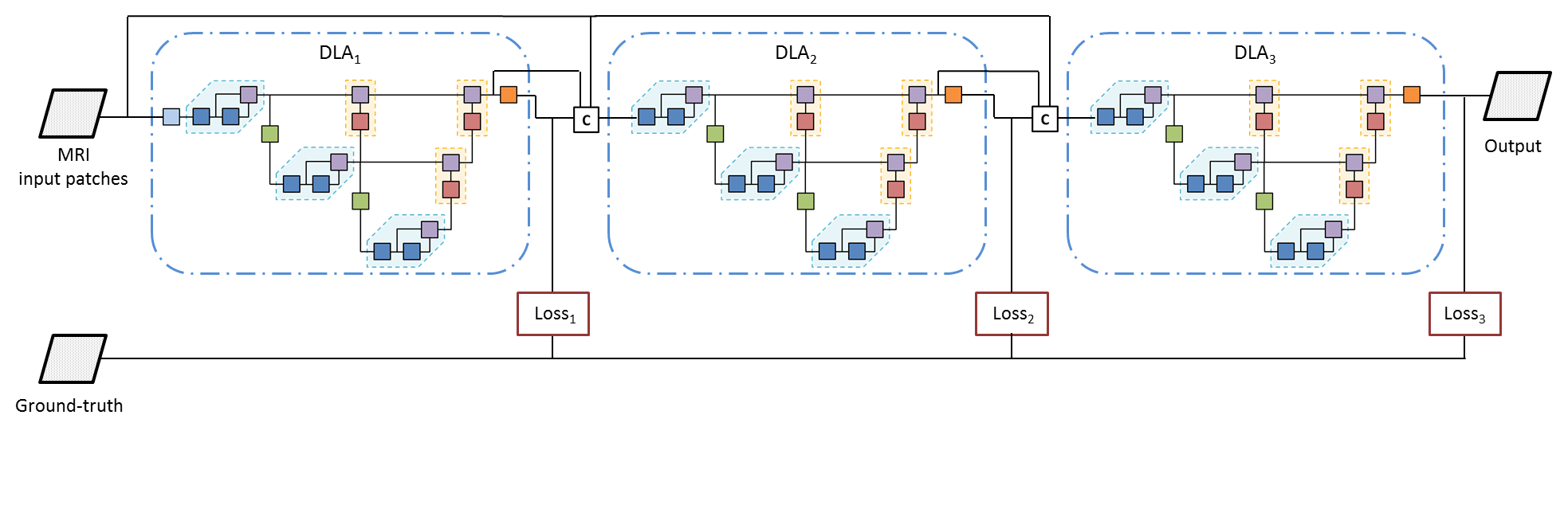}
		\centering
		\caption{Overview of the proposed Network. From one stage to another each module considers the MRI input patches alongside the respective feature maps of the last previous layer and the probability maps from the Softmax.}
		\label{fig:Overall_Architecture}
\end{figure}

In this work, we address the problem of brain tumor segmentation using a 2D FCN. We propose a variant of the DLA architecture \cite{yu2018deep} as the core segmentation architecture. An overview of the proposed architecture is depicted in Fig. \ref{fig:Overall_Architecture}.

The first stage receives as input multi-sequence MRI images of size $120 \times 120$, generating a first rough segmentation of the brain tumor. The subsequent stages of the architecture receive the probability map and also the feature maps from the previous stage. These are combined with the raw input images to provide a more accurate segmentation of the brain tumor. The complete neural network is trained end-to-end, having a loss function for each stage.

\subsection{Deep Layer Aggregation}
    Widely known deep learning-based architectures, e.g. U-Net \cite{ronneberger2015u} and FCN \cite{shelhamer2017fully} consider the information from shallow layers by employing linear skip connections. However, this linear aggregation, \ie the combination of different blocks of a network, restrains the possibility to refine features from shallow stages of the network. Deep layer aggregation (DLA) \cite{yu2018deep} extends over linear aggregation layers to better fuse across channels and depths (semantic fusion), and across resolutions and scales (spatial fusion). Considering more depth and sharing of features extracted from different stages of the network improves the overall inference.
    
    Deep Layer Aggregation considers two deep aggregation blocks: Iterative Deep Aggregation (IDA) and Hierarchical Deep Aggregation (HDA). IDA focus on spatial fusion, increasing the spatial resolution while simultaneously elaborating over the input. HDA aims for semantic fusion, extending linear skip connections with a tree-based structure that spans the feature hierarchy over different depths. HDA preserves feature channels while, at the same time, combines them with the feature channels of the current depth \cite{yu2018deep}.
    
    Having the capability to better aggregate features from shallow levels of the network and further on refine them, poses as an interesting approach for brain tumor segmentation. With semantic fusion we aim to properly distinguish healthy from glioma tissue, while with spatial fusion we aim to properly locate it and distinguish the different types of glioma tissue.
    
    Down-sampling is usually achieved through pooling or convolutional layers with stride of 2. However, these procedures may lead to aliasing that may impact the details of the segmentation \cite{zhang2019making}. Instead, we employ Max-pooling with kernel size of 2 and stride of 1, followed by Gaussian filtering with stride 2 and kernel size of $5 \times 5$ and a sigma of $1.25$. As for the Up-sampling block, we employed Transpose Convolutions with a kernel size of $3\times3$ and a stride of $2\times2$, as opposed to typical up-sampling layers.
    
\begin{figure}[!h]
			\includegraphics[width=\textwidth]{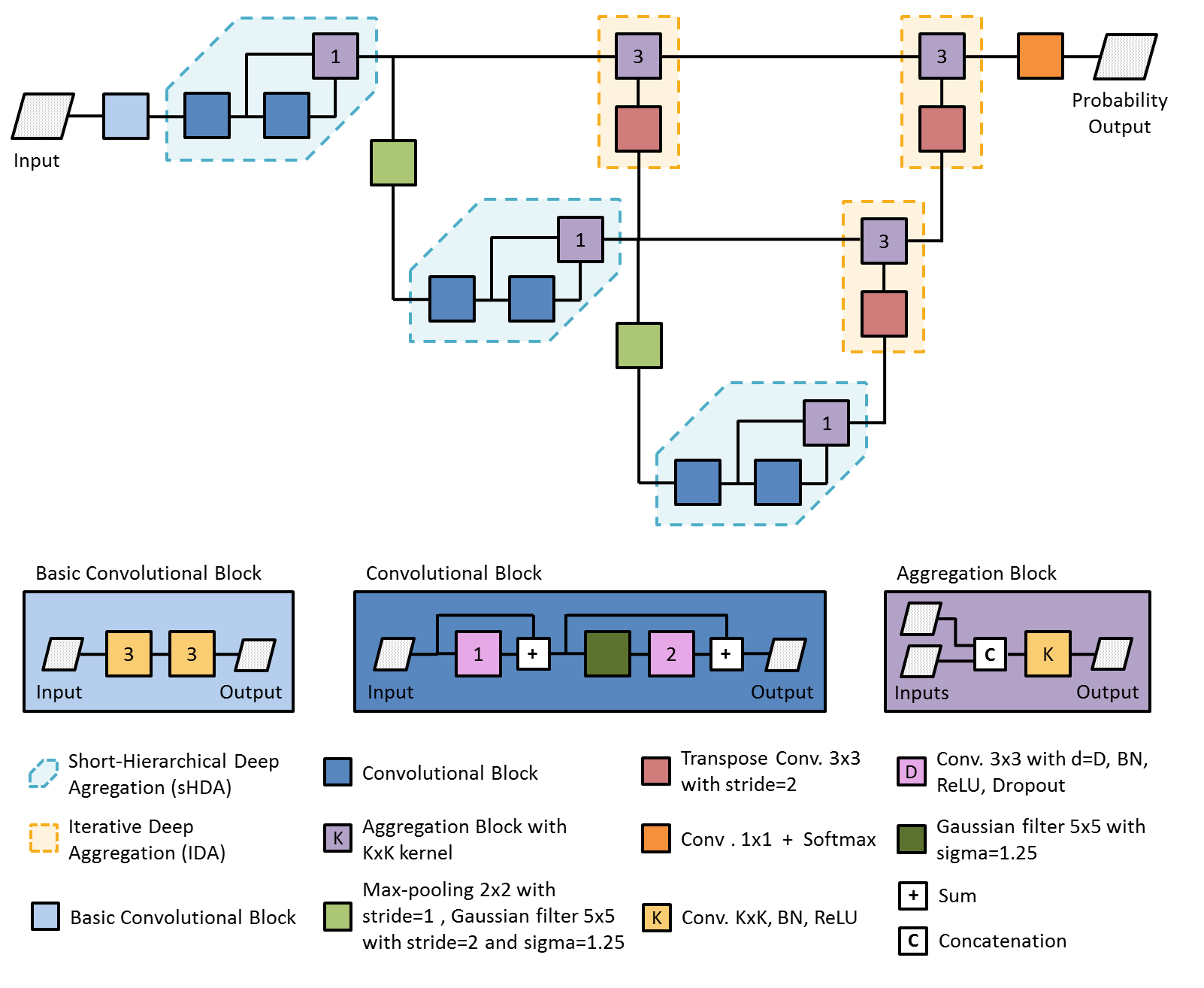}
			\centering
			\caption{Deep Layer Aggregation module, employed at different levels of our proposed architecture.}
			\label{fig:DLA_Network}
	\end{figure}

    The DLA architecture is presented in Fig. \ref{fig:DLA_Network}, detailing as well the employed operational block.

    \subsection{Loss function}
    
   The proposed architecture was trained with three auxiliary losses, one for each level. The equation of  the complete loss is:

    \begin{equation}
        L = 0.3\times\loss_1 + 0.4\times\loss_2 + 1.0\times\loss_3,
    \end{equation}
    
    Each auxiliary loss considers the categorical cross entropy, defined as:
    
    \begin{equation}
        \mathcal{L}(y, \hat{y}) = - \sum_i y_i log(\hat{y}_i),
    \end{equation}
    
    \noindent where $i$ indexes the class, $\hat{y}$ the probabilistic prediction and $y$ the true class, when employing one-hot encoding. This loss function was also computed over brain tumorous region \cite{isensee2018no,zhao2019bag}.
    
\section{Experimental Setup}
\label{sec:setup}

\subsection{Data}
The proposed methods were evaluated in the BraTS 2020 dataset \cite{bakas2017segmentationa,bakas2017segmentationb,bakas2017advancing,bakas2018identifying,menze2014multimodal}, which consists of multi-institutional MRI acquisitions with both HGGs and LGGs. Additionally, the acquired MRI sequences follow the clinical practice, hence including 4 sequences: T1, contrast enhanced T1, T2, and FLAIR. The Training set includes $369$ subjects with the corresponding manual segmentations. For development, we randomly split the images into training ($295$ patients), validation ($37$ patients), and test (37 patients). BRATS 2020 also includes an official Validation set with $125$ subjects, where the manual annotations are not publicly available. Therefore, segmentation scores are blindly computed by an online platform.

\subsection{Pre-processing and Data Augmentation}

The dataset was already pre-processed, namely multi-sequence registration and skull stripping. Hence, we only performed intensity normalization by assuring that each patient brain volume had zero mean and unitary standard deviation, using its own statistics.

The data augmentation used, to further prevent overfitting and improve the generalization capacity of the proposal, encompasses random shifts ($-0.1$\,--\,$0.1$) and scale ($0.9$\,--\,$1.1$) to each input channel after patch standardization. Also, we apply bi-dimensional stochastic rotation and cutout \cite{taylor2017}.

\subsection{Settings and model training}

Due to memory constraints, we employ a 2D axial patch-wise approach, with a patch size of $120 \times 120$. The training patches are sampled such that the tumor is partially covered. The model is trained end-to-end by backpropagation. The loss function is optimized using the AdamW \cite{loshchilov2017decoupled} optimizer. For regularization purposes, we employ weight decay. It is kept constant at $\num{1e-3}$ for the first $50$ epochs, then it decreases using a cosine annealing to $\num{1e-6}$. A similar scheduling is applied to the learning rate; however, it is initialized at $\num{1e-4}$ and decays till $\num{5e-5}$. We also employ spatial dropout \cite{tompson2015efficient} with probability of $0.25$.

The neural network was trained for $170$ epochs. For the choice of the neural network structure and hyperparameters, we used the validation and test sets for selecting the best epoch and to compare different set-ups and choose the final neural network, respectively. These validation and test sets were obtained from the training data.

For training the ensemble we split the training dataset in five folds. For each fold, we split it in two, using half for validation and the other half for testing. The remaining four folds are used for training the neural network. In each fold, we changed the seed used for initializing the weights of the neural network. The initialization of the weights condition the region of the loss where we start the optimization. So by changing the seed, we aim to obtain diverse solutions, which we know that may improve the performance of the ensemble.

For evaluation, we report two metrics: Dice Score, and Hausdorff Distance \cite{bakas2017advancing,menze2014multimodal}. The proposed methods were implemented using PyTorch, and the experiments were conducted on an Nvidia RTX 2080 TI GPU.

\subsection{Post-processing}
    
    As post-processing step, we employed morphological filtering to remove small clusters whose volume is less than a threshold. The threshold was chosen based on statistics obtained on the training dataset.
    
    \begin{figure}[!h]
			\includegraphics[width=12cm]{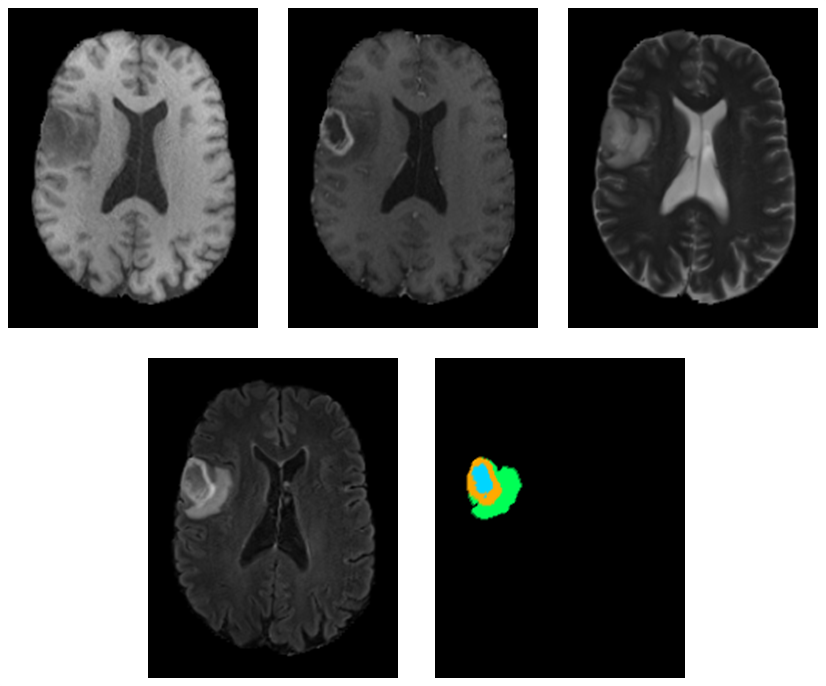}
			\centering
			\caption{Qualitative results from patient BraTS20\_Validation\_002. The enhancing tumor is shown in orange, necrosis in turquoise and edema in green. This subject obtained a mean Dice score of $0.9272$, $0.9603$ and $0.8939$ for whole tumor, core tumor and enhanced tumor, respectively.}
			\label{fig:seg-a}
	\end{figure}
	
	\begin{figure}[!h]
			\includegraphics[width=12cm]{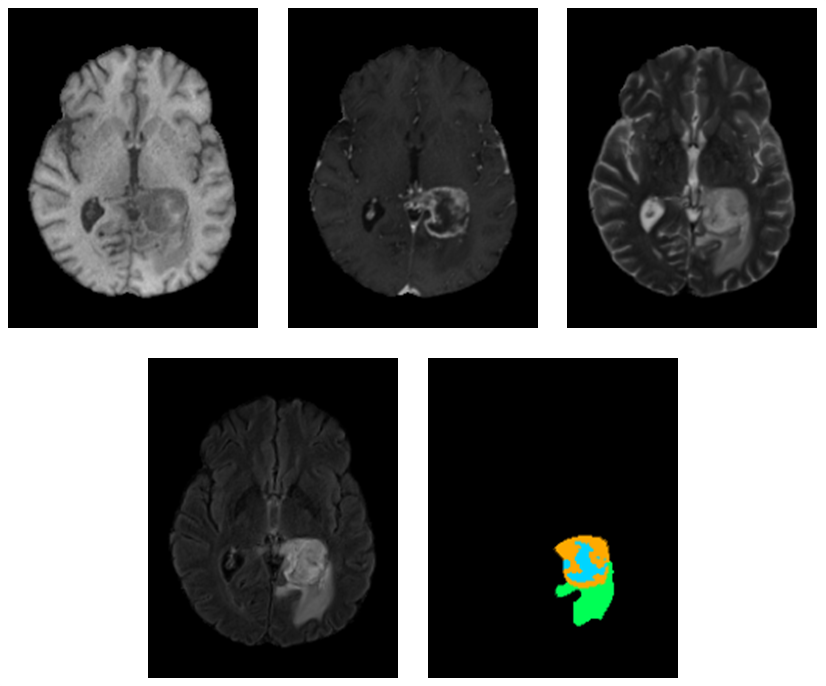}
			\centering
			\caption{Qualitative results from patient BraTS20\_Validation\_017. The enhancing tumor is shown in orange, necrosis in turquoise and edema in green. This subject obtained a mean Dice score of $0.9500$, $0.9354$ and $0.9446$ for whole tumor, core tumor and enhanced tumor, respectively.}
			\label{fig:seg-b}
	\end{figure}

\section{Results}
\label{sec:results}

\newcolumntype{C}{>{\centering\arraybackslash}X}

\begin{table}
\caption{Dice (DSC) and Hausdorff (HD\textsubscript{95}) metrics of the proposed method on BraTS 2020 Validation set. Each metric is represented by the average. WT - whole tumor, TC - tumor core, ET - enhancing tumor core.}
\begin{tabularx}{\textwidth}{@{} lCCCCCC @{}}
 % Header
\toprule
\multirow{2.5}{*}{} 
& \multicolumn{3}{c@{}}{DSC}
& \multicolumn{3}{c@{}}{HD\textsubscript{95}}\\
    % Subheader
    \cmidrule(l){2-4}
    \cmidrule(l){5-7}
    & ET & WT & TC & ET & WT & TC \\
% Results
\midrule
%%% Method 1
DLA-Pool & 0.7246 & 0.8916 & 0.7885 & 33.87 & 8.06 & 8.93 \\
\addlinespace

DLA-GConv & 0.7418 & 0.8918 & 0.7945 & 27.81 & 7.07 & 12.99 \\
\addlinespace

Ensemble & 0.7565 & 0.9050 & 0.8062 & 27.16 & 4.34 & 9.39 \\

    \bottomrule
    \end{tabularx}
\label{tab:brats2020_val}
\end{table}

We report our results on BRATS 2020. In table \ref{tab:brats2020_val}, we present the results on BRATS 2020 validation dataset, which were computed using the online platform.

The first entry in the table is DLA-Pool neural network. This uses a $2 \times 2$ max-pooling layer with stride of $2$ to increase the field of view, which is a common approach. In DLA-GConv, we evaluate the observation by Zhang \cite{zhang2019making} that max-pooling may cause aliasing, so we employ a max-pooling with kernel size of 2 and stride of 1, followed by a Gaussian filtering with stride 2 and kernel size of $5 \times 5$ and a sigma of $1.25$. The effect of the Gaussian filter is to limit the rapid variation on the input pattern, avoiding aliasing. As can be observed in Table \ref{tab:brats2020_val}, this filtering in the downsampling improved the segmentation in all regions, specially the core and enhancing tumor. In this case, all metrics improved, except for the Hausdorff distance of the core tumor. The improvement was more pronounced in the enhancing tumor. In the last entry, we present the results of our ensemble. As can be observed the ensemble improved all metrics.

In Figures \ref{fig:seg-a} and \ref{fig:seg-b}, we show two example of the segmentation obtained with our method in the BraTS 2020 Validation set using the ensemble.

In Table \ref{tab:brats2020_test}, we present the results obtained in the BraTS 2020 Test set. We observe that there was a large drop in the whole tumor, but the core tumor and enhancing tumor increased when comparing to the results on the BraTS 2020 Validation set. Also, the Haudorff distance improved, except for the core tumor. Comparing the mean Dice with the Dice value in the interquartile, we also note that the mean Dice are closer to the lower interquartile value and in the core tumor it is even lower than the interquartile value. This may indicate that we have some low outliers that are skewing the mean Dice. The identification of these cases may help to spot aspects to improve in our approach, which we leave as future work.

\newcolumntype{C}{>{\centering\arraybackslash}X}

\begin{table}
\caption{Dice (DSC) and Hausdorff (HD\textsubscript{95}) metrics of the proposed method on BraTS 2020 Test set. WT - whole tumor, TC - tumor core, ET - enhancing tumor core.}
\begin{tabularx}{\textwidth}{@{} lCCCCCC @{}}
 % Header
\toprule
\multirow{2.5}{*}{} 
& \multicolumn{3}{c@{}}{DSC}
& \multicolumn{3}{c@{}}{HD\textsubscript{95}}\\
    % Subheader
    \cmidrule(l){2-4}
    \cmidrule(l){5-7}
    & ET & WT & TC & ET & WT & TC \\
% Results
\midrule
%%% Method 1
Mean & 0.7900 & 0.8858 & 0.8297 & 20.44 & 5.32 & 22.32 \\
\addlinespace

StdDev & 0.2278 & 0.1175 & 0.2514 & 79.73 & 7.60 & 79.42 \\
\addlinespace

Median & 0.8514 & 0.9208 & 0.9187 & 1.41 & 3.00 & 2.24 \\
\addlinespace

25 quantile & 0.7698 & 0.8786 & 0.8624 & 1.00 & 2.00 & 1.41 \\
\addlinespace

75 quantile & 0.9181 & 0.9478 & 0.9567 & 2.40 & 5.00 & 5.38 \\
\addlinespace

    \bottomrule
    \end{tabularx}
\label{tab:brats2020_test}
\end{table}

\section{Conclusion}
\label{sec:conclusion}

The segmentation of gliomas in MRI is a challenging task, due to the heterogeneity of the lesion and the imaging modality itself. In this entry to the BraTS 2020 challenge, we propose a multi-cascade FCN for brain tumor segmentation based on the Deep Layer Aggregation principles, which consists in a systematic aggregation of features from the different scales. Also, we investigated the use of Gaussian filters for reducing the aliasing during pooling, which we found to be beneficial.

Our neural network was able to obtain a Dice score of $0.8858$, $0.8297$ and $0.7900$, with an Hausdorff Distance of $5.32$ mm, $22.32$ mm and $20.44$ mm for the whole tumor, core tumor and enhanced tumor, respectively, in the BraTS 2020 Test set.

%%%%%\newpage
%%%%% Table of RBM Ablative Studies %%%%%

% Use the command \cite only!
\bibliographystyle{splncs04}
\bibliography{references}

\end{document}